\documentclass[english,twocolumn,pre,aps]{revtex4}
\usepackage[T1]{fontenc}
\usepackage[latin9]{inputenc}
\usepackage{graphicx}
\usepackage{amssymb}

\makeatletter

\usepackage{babel}
\makeatother

\begin{document}

\title{Homopolar oscillating-disc dynamo driven by parametric resonance}

\author{J\={a}nis Priede}

\affiliation{Applied Mathematics Research Centre, Coventry University, Coventry,
CV1 5FB, United Kingdom }

\author{Ra\'ul Avalos-Z\'u\~niga}

\affiliation{Universidad Aut\'onoma Metropolitana, D.F., M\'exico}

\author{Franck Plunian}

\affiliation{Universit\'e Joseph Fourier, CNRS, LGIT, Grenoble, France}

\begin{abstract}
We use a simple model of Bullard-type disc dynamo, in which the disc
rotation rate is subject to harmonic oscillations, to analyze the
generation of magnetic field by the parametric resonance mechanism.
The problem is governed by a damped Mathieu equation. The Floquet
exponents, which define the magnetic field growth rates, are calculated
depending on the amplitude and frequency of the oscillations. Firstly,
we show that the dynamo can be excited at significantly subcritical
disc rotation rates when the latter is subject to harmonic oscillations
with a certain frequency. Secondly, at supercritical mean rotation
rates, the dynamo can also be suppressed but only in narrow frequency
bands and at sufficiently large oscillation amplitudes.
\end{abstract}
\maketitle
In dynamo experiments, a high driving power is necessary to achieve
the self-excitation of the magnetic field. The ensuing liquid metal
flow is usually strongly turbulent. In both Riga and Karlsruhe dynamo
experiments \cite{Gailitis00-1,Stieglitz01}, the turbulent fluctuations
were partly inhibited by the internal walls, whereas in the Cadarache
experiment \cite{Monchaux07}, the absence of such walls resulted
in large-scale flow fluctuations \cite{delaTorre07}. The effect
of flow fluctuations on the dynamo threshold has been addressed in
several recent studies \cite{Fauve03,Normand03,Leprovost04,Leprovost05,Laval06,Petrelis06,Volk06,Peyrot07,Peyrot08,Muller09}.
Solving the kinematic dynamo problem for a given non-stationary flow
usually governed by the Navier-Stokes equations shows that turbulence
generally has an adverse effect on the dynamo excitation, unless the
fluctuations are strong enough to drive the dynamo by themselves without
any mean flow. In the latter case, we speak of a fluctuation dynamo
\cite{Schekochihin07,Stepanov08}, whose experimental
implementation seems hardly feasible because of the high excitation
threshold. However the possibility that fluctuations excite the magnetic
field by the parametric resonance mechanism \cite{Landau-69} can
not be excluded. Parametric resonance has been proposed in the somewhat
different context of spiral galaxies as promoter of bisymmetric
magnetic field structure \cite{Chiba90,Schmitt92,Kuzanyan93,Moss96}.

In this paper, we use a simple model of the Bullard-type disc dynamo
\cite{Bullard-1955} to show that the magnetic field can indeed be
excited by the parametric resonance mechanism, even when relatively small
harmonic oscillations are added to significantly subcritical disc
rotation rates.

\begin{figure}
\begin{centering}
\includegraphics[width=0.45\textwidth]{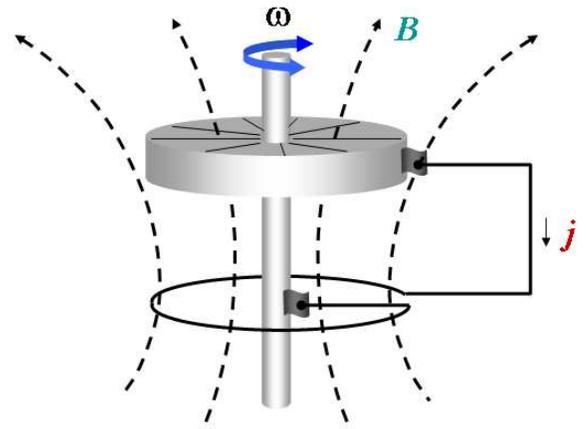} 
\par\end{centering}

\caption{\label{fig:Sketch}Sketch of a homopolar disc dynamo.}

\end{figure}

\begin{figure*}
\begin{centering}
\includegraphics[width=0.45\textwidth]{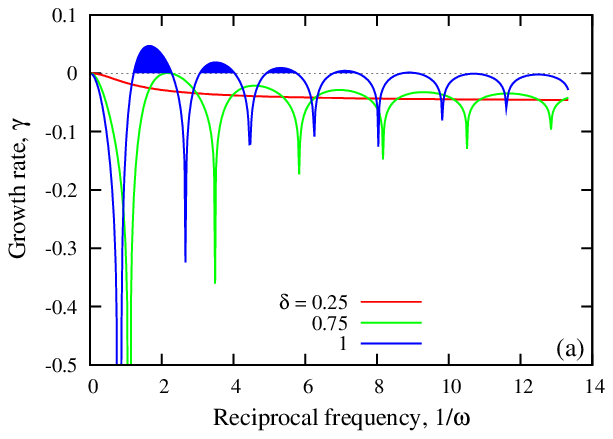}~\includegraphics[width=0.45\textwidth]{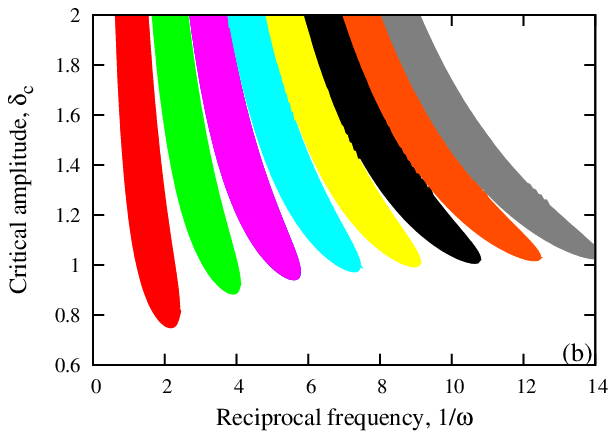} 
\par\end{centering}

\caption{\label{fig:a0}Growth rate $\gamma$ versus the reciprocal frequency
(a) and the critical amplitude $\delta_{c}$ versus the frequency
(b) for the marginal mean rotation rate at $\alpha_{0}=0.$}

\end{figure*}

Consider a Bullard-type disc dynamo \cite{Bullard-1955} which consists
of a solid conducting disc rotating with a generally time-dependent
angular velocity $z(t)$ about its axis, and a wire twisted around
the axle and connected by sliding contacts to the rim of the disc
and the axle as shown in Fig. \ref{fig:Sketch}. The disc is assumed
to be segmented so that azimuthal current can flow only at its rim.
This corresponds to the modification of the Bullard disc dynamo suggested
by Moffatt in order to eliminate exponential growth of the magnetic
field in the limit of a perfectly conducting disc \cite{Moffatt-1979}.
The system is described by the following set of dimensionless equations
(for details see Ref. \cite{Moffatt-1979})\begin{equation}
\begin{array}{ccl}
\dot{x} & = & r(y-x),\\
\dot{y} & = & xz+mx-(m+1)y,\\
\dot{z} & = & g\left[1+x(mx-(m+1)y)\right]-kz,\end{array}\label{eq:xyz}\end{equation}
 where $x$ and $y$ are magnetic fluxes through the loop made by
the wire and the rim of disc, respectively; $z$ is the dimensionless
angular velocity of the disc; $r$ accounts for the resistance of
the disc relative to that of the loop, and $m$ characterizes the
relative mutual inductance of the disc and the loop; the dot stands
for the time-derivative $d/dt.$ The disc is driven by a generally
time-dependent torque $g,$ and braked by a viscous-type friction
characterized by the coefficient $k$ which is necessary for the structural
stability of the system \cite{Hide-1995}. Henceforth, we assume
the friction to be strong with respect to the inertia of the disc
accounted for by $\dot{z}$ in Eq. (\ref{eq:xyz}), which, thus, results
in $z=z_{0}\left[1+x(mx-(m+1)y)\right],$ where $z_{0}=g/k$. The
remaining two 1st-order ODEs in (\ref{eq:xyz}) can be combined into
a single 2nd-order Duffing-type equation \cite{Leprovost04} with
a non-linear friction\begin{equation}
\ddot{x}+(1+\beta x^{2})\dot{x}-\alpha x+\lambda x^{3}=0,\label{eq:x}\end{equation}
 where $x$ and $t$ are rescaled by $(m+1+r)$ and $(m+1+r)^{-1}$,
respectively, and $\alpha=r(z_{0}-1)/(m+1+r)^{2},$ $\beta=z_{0}(m+1)(m+1+r),$
and $\lambda=rz_{0}.$ Further, we focus on the evolution of small
initial perturbations of the magnetic field characterized by $x\ll1,$
for which Eq. (\ref{eq:x}) can linearized by setting $\beta=\lambda=0.$
Then the only remaining parameter $\alpha$ depends directly on the
deviation of the disc rotation rate from its critical value $\alpha=0$.
For $\alpha>0$, a small initial magnetic field starts to grow exponentially
provided that the disc rotates steadily \cite{Bullard-1955}. In
this study, we are interested in how the generation of the magnetic
field is affected by the unsteadiness of the disc rotation \begin{equation}
\alpha=\alpha_{0}+\delta\cos(\omega t),\label{eq:alpha01}\end{equation}
which besides the mean part $\alpha_{0}$ contains also an oscillatory
component with the amplitude $\delta$ and the circular frequency
$\omega.$ Then the linearized Eq. (\ref{eq:x}) reduces to a damped
Mathieu equation \[
\ddot{x}+\dot{x}-(\alpha_{0}+\delta\cos(\omega t))x=0.\]
 Using the substitution $x(t)=\exp(-t/2)\chi(\omega t/2),$ the equation
above can be transformed into the canonical Mathieu equation\begin{equation}
\ddot{\chi}+\left[a-2q\cos(2\tau)\right]\chi=0,\label{eq:matthieu}\end{equation}
 where $a=-(1+4\alpha_{0})/\omega^{2}$, $q=2\delta/\omega^{2}$ and
$\tau=\omega t/2$. According to Floquet theory, a particular solution
to Eq. (\ref{eq:matthieu}) can be written as $\chi(\tau)=\exp(i\nu\tau)f(\tau),$
where $f(\tau)$ is a $\pi$-periodic function and $\nu$ is the Floquet
exponent---both dependent on the parameters $a$ and $q$. According
to this solution, the amplitude of the magnetic field $x(t)$ evolves
exponentially in time with the maximum growth rate $\gamma=\left(\left|\Im\left[\omega\nu\right]\right|-1\right)/2,$
where the modulus accounts for the time-reflection symmetry of Eq.
(\ref{eq:matthieu}) \cite{Bender-Orsag-1978}. Thus, the amplitude
of the magnetic field grows exponentially when $\gamma>0,$ whilst
the marginal state is defined by $\gamma=0.$ We use the Maple computer
algebra software to calculate Floquet exponent which defines the growth
rate $\gamma$ depending on the amplitude $\delta$ and the frequency
$\omega.$ Next, we find the critical oscillation amplitude $\delta_{c}$
as the function of frequency $\omega$ for fixed values of the mean
rotation $\alpha_{0}$ by solving numerically equation $\left|\Im\left[\omega\nu\right]\right|=1$
corresponding to $\gamma=0.$ Eventually, we determine the minimal
oscillation amplitude $\delta_{\min}$ and the corresponding frequency
at which an exponentially growing magnetic field first appears. The
corresponding numerical results are presented and discussed below.

\begin{figure*}
\begin{centering}
\includegraphics[width=0.45\textwidth]{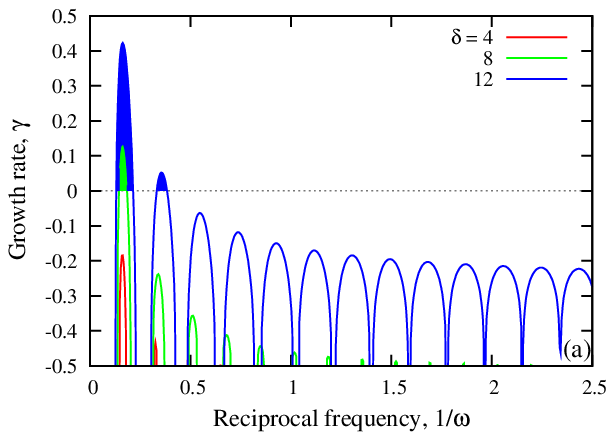}~\includegraphics[width=0.45\textwidth]{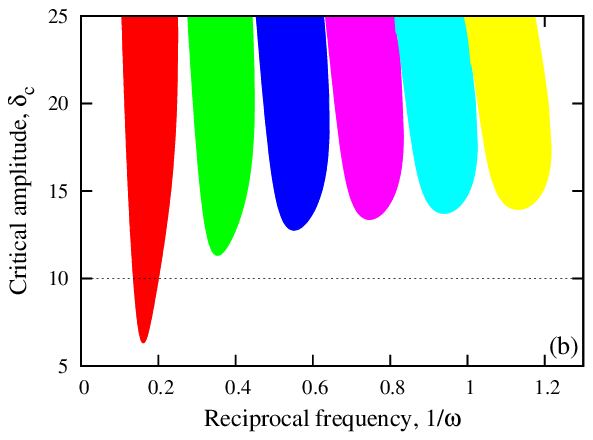} 
\par\end{centering}

\caption{\label{fig:a-10}Growth rate $\gamma$ versus the reciprocal frequency
(a) and the critical amplitude $\delta_{c}$ versus the frequency
for a strongly subcritical mean rotation rate at $\alpha_{0}=-10.$ }

\end{figure*}

\begin{figure}[b]
\begin{centering}
\includegraphics[width=0.45\textwidth]{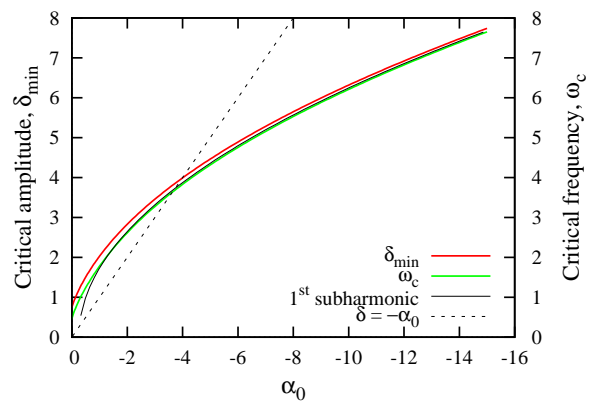} 
\par\end{centering}

\caption{\label{fig:dltalph}Minimal oscillation amplitude and critical frequency
versus subcritical rotation rate parameter $(\alpha<0).$ }

\end{figure}

We start with a marginal mean disc rotation rate $\alpha_{0}=0,$
which corresponds to the dynamo excitation threshold when the disc
rotates steadily, i.e., $\gamma=0$ when $\delta=0.$ As seen in Fig.
\ref{fig:a0}(a), the disc oscillations about the critical rotation
rate with a sufficiently small amplitude $(\delta=0.25)$ brings the
growth rate to a constant level below zero as the frequency is reduced
(reciprocal frequency increased). As the oscillation amplitude increases,
first, the growth rate splits into separate frequency bands whose
width decreases as $\sim1/\omega$ for $\omega\rightarrow0.$ In order
to show this increasingly fine-scale structure of the growth rate
as $\omega\rightarrow0$, we use the reciprocal frequency which is
proportional to the period of disc oscillations. Secondly, the growth
starts to increase with the oscillation amplitude and approaches zero
again at $\delta\approx0.75$ for a certain critical frequency. Further
increase in the oscillation amplitude to $\delta=1$ results in the
appearance of several frequency bands with positive growth rates $(\gamma>0)$
which are shown as filled regions in Fig. \ref{fig:a0}(a). The critical
amplitude $\delta_{c}$ at which the growth rate turns zero, is shown
for $\alpha_{0}=0$ in \ref{fig:a0}(b) against the oscillation frequency.
Marginal state with $\gamma=0$ corresponds to the boundaries of the
separate frequency bands shown by different colors. Growth rate is
positive corresponding to the dynamo action inside the filled frequency
bands which approach each other closely as the oscillation amplitude
increases. Note that only a certain number of first instability bands
are shown in Fig. \ref{fig:a0}(b). There is an infinite sequence
of similar instability bands of decreasing width as $\omega\rightarrow0.$
Thus, the range of unstable frequencies, which is intervened by infinitely
many, much narrower stability bands, extends down to $\omega=0,$
whereas it is bounded from above by the first instability band. The
critical oscillation amplitude, which is the lowest for the first
instability band, rises to an asymptotic value depending on $\alpha$
as $\omega\rightarrow0.$ The minimal value of the critical oscillation
amplitude and the corresponding frequency at which it occurs are plotted
in Fig. \ref{fig:dltalph} versus subcritical (negative) values of
$\alpha.$

The reduction of disc rotation rate to a moderately subcritical value
of $\alpha_{0}=-1,$ results in the increase of the minimal oscillation
amplitude necessary for the dynamo action $(\gamma>0)$ up to $\delta_{c}\approx2.$
It means that the maximum disc rotation rate (\ref{eq:alpha01}) during
the oscillation cycle temporally exceeds the critical value $\alpha=0$
for a steady rotation. This, however, changes when the disc rotation
rate is reduced further down to $\alpha_{0}=-10,$ for which the growth
rates and the corresponding generation bands are shown in Fig. \ref{fig:a-10}.
In this case, a positive growth rate band is seen in Fig. \ref{fig:a-10}(a)
already at $\delta_{c}=8.$ This implies that $\alpha<0$ during the
whole cycle of disc oscillation. Thus, the dynamo appears at a maximum
disc rotation rate below its critical value for a steady rotation.
There is a range of frequencies seen Fig. \ref{fig:a-10}(b), at which
a subcritical growth of the magnetic field is possible with the the
oscillation amplitude $\delta_{c}<-\alpha_{0}=10.$ As seen in Fig.
\ref{fig:dltalph}, such a subcritical excitation of magnetic field
appears already at $\alpha_{0}\lesssim-4,$ where the minimal required
oscillation amplitude $\delta_{\min}$ becomes smaller than $-\alpha_{c},$
which is shown by the dashed line. Note that for a strongly subcritical
$\alpha_{0}$ and steadily rotating disc, an initial perturbation
of the magnetic field oscillates with the period $O(1/\sqrt{-\alpha_{0}})$
which is much shorter than the characteristic damping time $O(1)$.
Thus, for such $\alpha_{0}$ the damping of magnetic field oscillations
becomes relatively slow that enables parametric excitation of the
magnetic field by a relatively weak modulation of the disc rotation
rate.

For $\alpha_{0}<-0.25,$ the bands of unstable frequencies are associated
with the subharmonics of Mathieu equation defined by $a^{1/2}=1,2,3,\dots$
in Eq. (\ref{eq:matthieu}) \cite{Bender-Orsag-1978}. Thus, the
most unstable band corresponds to the first subharmonic with $\omega_{1}=\sqrt{-4\alpha_{0}-1}$
which is seen in Fig. \ref{fig:dltalph} to approximate $\omega_{c}$
well for $\omega\gtrsim1.$ In addition, it is interesting to note
that the critical frequency, $\omega_{c}$, at which $\delta_{\min}$
occurs, is numerically very close to $\delta_{\min}$ itself in the
whole range of $\alpha_{0}<0.$

\begin{figure*}
\begin{centering}
\includegraphics[width=0.45\textwidth]{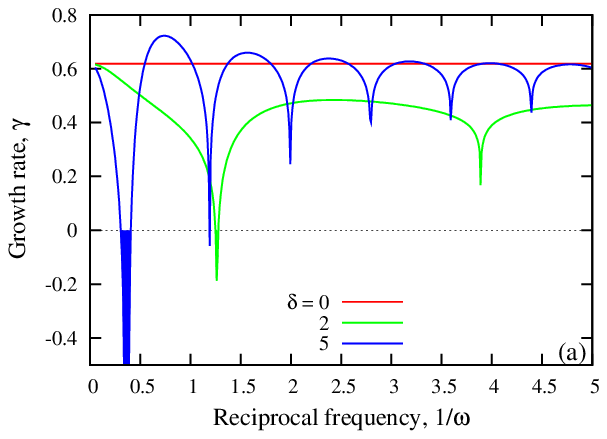}~\includegraphics[width=0.45\textwidth]{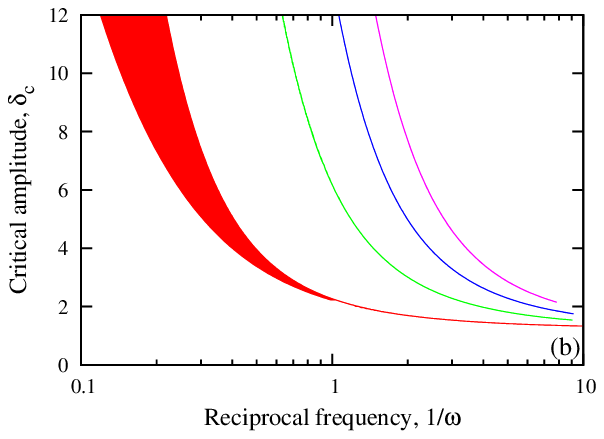} 
\par\end{centering}

\caption{\label{fig:a1}Growth rate $\gamma$ versus the reciprocal frequency
(a) and the critical modulation amplitude $\delta_{c}$ versus the
frequency for a slightly supercritical mean rotation rate at $\alpha_{0}=1.$
Filled curves correspond to suppressed dynamo.}

\end{figure*}

For a supercritical disc rotation rate with $\alpha_{0}=1,$ the growth
rate $\gamma,$ which is seen in Fig. \ref{fig:a1}(a) to be positive
for a steadily rotating disc $(\delta=0$), first reduces as the oscillation
amplitude is increased to $\delta=2.$ Moreover, $\gamma$ is seen
to become negative within certain, relatively narrow frequency bands.
These negative growth rate bands $(\gamma<0)$, where the dynamo is
suppressed by the disc oscillations, are shown in Fig. \ref{fig:a1}(a)
by filled curves. As seen in Fig. \ref{fig:a1}(b), the width of the
first suppressed frequency band noticeably increases while the whole
band shifts towards the high-frequency range, i.e., low reciprocal
frequencies, as the oscillation amplitude increases. At the same
time, the width of the subsequent frequency bands, where the dynamo
is suppressed, remains very small, while the spacing between the adjacent
suppression bands reduces in terms of the reciprocal frequency.

In conclusion, we have shown in this study that dynamo can be excited
by the parametric resonance at considerably low velocities of the conducting
medium when the latter is subject to harmonic oscillations. Even though
this was demonstrated for a simple model Bullard-type disc dynamo,
we expect that similar mechanism may also be relevant for more realistic
dynamos.

\end{document}